\documentclass{emulateapj}
\renewcommand{\epsilon}{\varepsilon}

\newcommand{\dF}{{^{^*}\!\!F}}
\newcommand{\bP}{{\bf P}}
\newcommand{\bF}{{\bf F}}
\newcommand{\bU}{{\bf U}}

\newcommand{\del}{{\partial}}

\shorttitle{GRB central engine}
\shortauthors{Agnieszka Janiuk}

\begin{document}

\title{Microphysics in the gamma ray burst central engine}

\author{Agnieszka Janiuk$^{1}$}
\affil{$^{1}$
Center for Theoretical Physics, Polish Academy of Sciences, 
Al. Lotnikow 32/46, 
02-668 Warsaw, Poland}
\email{agnes@cft.edu.pl}

\begin{abstract}
We calculate the structure and evolution of a gamma ray burst
central engine 
where an accreting
torus has formed around the newly born black hole.  
We study the general relativistic, MHD
models and we self-consistently incorporate the nuclear equation of state.  
The latter accounts for the degeneracy of relativistic electrons, protons, and neutrons, and is used in the dynamical simulation,
instead of a standard 
polytropic $\gamma$-law. The EOS provides the conditions
for the nuclear pressure in the function of density and temperature, which evolve with time according to the conservative MHD scheme.
We analyze the structure of the torus and outflowing winds, and compute the 
neutrino flux emitted through the nuclear reactions balance in the dense and hot matter. We also estimate the rate of 
transfer of the black hole rotational energy to the bipolar 
jets. 
Finally, we elaborate on the nucleosynthesis of heavy elements in the accretion flow and the wind, through computations of the thermonuclear reaction network.
We discuss the possible signatures of the radioactive elements decay in the accretion flow. 
We suggest that further detailed modeling of the accretion flow in GRB engine,
together with its microphysics,
 may be a valuable tool to constrain the black hole mass and spin.
It can be complementary to the gravitational wave analysis, if the waves are detected 
with an electromagnetic counterpart. 
\end{abstract}

\section{Introduction}

Gamma Ray Bursts (GRB),  are
extremely energetic transient events, visible from the most distant parts of
the Universe \citep{Piran04,Kumar15}. 
 When a newly born black hole forms, either via the core collapse supernova, or after the merger of two compact stars, a
large amount of matter is accreted onto it within the timescale between tens 
of milliseconds, to thousands of seconds, with hyper-Eddington rates \citep{woosley93,paczynski98}. 
In such conditions, the nuclear reactions inside the plasma lead to 
creation of neutrinos, mainly via the electron-positron capture on nucleons,
which acts in the so-called NDAF models \citep{Narayan00,Lei09,Cao14}. 
The annihilation of neutrino-antineutrino pairs is a source of power
to the relativistic jets launched along the axis of the disk-black hole system.
This can be comparable to the power extracted on the cost of the black hole spin, with the disk mediated by the magnetic fields. Both neutrinos and magnetic 
fields in the accreting matter can drive the wind outflow from the disk surface, and act as a collimation mechanism for the relativistic jets (see, e.g., 
\citealt{LeeRamirez2007} and references therein). 
The latter, are then the sites of the 
high-energy radiation produced at large distances from the engine, on the cost of the magnetic reconnection and conversion of the jet electromagnetic energy flux into heat (e.g., \citealt{Bromberg2016}).

Numerical computations of the structure and evolution of the
accretion flow in the gamma ray bursts engine begun with
steady-state, and time-dependent models, which were axially and vertically 
averaged, and based on the classical prescription for viscosity with the 
so-called $\alpha$-parameter
\citep{popham1999, dimatteo2002, kohri2002, kohri2005, reynoso2006, chen, janiuk2004, janiuk2007, janiuk2010}.
More recently, these accretion flows are being described by fully relativistic, 
MHD computations \citep{Shibata2007, nagataki09, barkovkomis2010, barkov2011, rezzola2011, janiuk13, kyuchi2015}.

In the outer parts of the disc, the heavy elements can be formed, 
and then the isotopes are prone to the radioactive decay \citep{Fujimoto2004}. 
This 
may in the future bring observable effects on the measured hard 
X-ray spectra of GRBs \citep{aj2014} as well as the additional peaks in the blue and infrared bands a few days after the prompt GRB emission \citep{martin2015}. 
In addition, the
gamma ray bursts contribute in this way to the galactic chemical evolution 
\citep{surman2014}.
The physical conditions in this engine, its density, temperature, and electron 
fraction, depend on the global parameters of the flow, such as the amount of matter and accretion rate, black hole mass and its spin, and the magnetic field. 
These conditions in turn will affect the total abundances of heavier 
elements synthesized within the flow. Therefore, a possible detection of the signatures 
of the radioactive decay of these species, and quantitative modeling of nucleosynthesis together with the central engine structure and evolution, and the jets gamma ray fluence, could give a constraint on the black hole parameters. These, in turn, can now be independently estimated, if the black hole is born via a merger event, detectable in the 
gravitational wave interferometer.

In this work, we study the central engine of a gamma ray burst, which is 
composed of a stellar mass,
rotating black hole and accreting torus that has formed from the remnant
matter at the base of the GRB jet. 
We compute the time-dependent, two dimensional model of the 
rotationally supported, magnetized disc, rapidly accreting onto the center.
The simulation is based on the HARM (High Accuracy Magnetohydrodynamics) scheme, which works on the stationary metric around black hole. The code integrates the total energy equation and updates the set of 'conserved' variables, i.e. comoving density, energy-momentum, and magnetic field \citep{gammie}. 

The basic version of the HARM scheme was designed to model the 
accretion flows in the centers of galaxies, where the gas equation of state is
with a good accuracy given by that of an ideal gas, and can be described using an adiabatic relation (e.g., \citealt{mmosc2013}).
In our current
dynamical model, we use the non-ideal equation of state, computed 
on the basis of the $\beta$-equilibrium \citep{janiuk2007}. We allow for the partial degeneracy of relativistic nucleons, electrons and positrons in the 
accreting plasma, and we compute the neutrino cooling rate due to the weak interaction processes, electron-positron pair anihillation, as well as bremsstrahlung and plasmon decay. 
Here,
we for the first time 
incorporate this detailed microphysics into the heart of the GR 
MHD scheme. The neutrino cooling was computed already from the nuclear reaction balance in the context of gamma ray bursts in 
our previous work \citep{janiuk13}. However, in that work, still the simple, adiabatic relation for the pressure ($p=(\gamma-1)u$, with $\gamma=4/3$) was used, and only the internal energy of the gas was updated during the simulation.
In the current model, we mainly address the question of microphysics in the accretion flow, and we are now also able to self-consistently estimate the efficiency of the disk neutrino cooling. We check if it is capable of powering the relativistic jets in
the gamma ray bursts, now in the case of a rather weak magnetic field. 
We also check the effect of the rotation speed of the black hole. In addition, we analyze the structure, velocity and amount of mass loss through the uncollimated winds launched from the 
accretion disc.
Finally, we examine the resulting nucleosynthesis in the body of the disk, 
which can occur already at small distances from the black hole, as well as in the winds,
and we show that the accretion flow 
can be the site of significant production of heavy element.

The article is organized as follows. In \S~\ref{sec:model}, 
we describe our MHD model of the GRB central engine. 
In \S~\ref{sec:results}, we present the results, describing the
structure of the torus and the effects of adopted microphysics treatment. 
We also compute the power transferred to the GRB jets via the 
neutrino anihillation and compare it to the
energy extracted by the magnetic fields dragged through the BH 
horizon.
In \S~\ref{sec:nucleo} we describe the process of heavy elements formation
in the outskirts of the torus in the GRB engine. We present the resulting abundances of elements computed under the assumption of nuclear statistical equilibrium.
Finally, we compare our results with the previous simulations.  
We discuss the results in \S~\ref{sec:diss}.

\section{Model of the hyperaccreting disk}\label{sec:model}
	
The model computations are based on the axisymmetric, general relativistic
MHD code {\it HARM-2D}, described by \citet{gammie} and \citet{noble}. 
It uses a conservative, shock-capturing 
scheme, and provides solver for the continuity and energy-momentum conservation equations, assuming a force-free approximation.
Here we use our own version of this code, which we supplemented with the 
numerical routines to
compute the equation of state and rate of cooling by neutrinos.

\subsection{Equation of state and neutrino cooling}

The neutrino cooling processes adopted in our computations are
the reactions of electron and positron capture on nucleons, the electron-positron pair anihillation, nucleon bremsstrahlung and plasmon decay.
The capture reactions are:
\begin{eqnarray}
\label{eq:urca}
p + e^{-} \to n + \nu_{\rm e} \nonumber \\
p + \bar\nu_{\rm e} \to n + e^{+}  \nonumber \\
p + e^{-} + \bar\nu_{e} \to n \nonumber \\
n + e^{+} \to p + \bar\nu_{\rm e} \nonumber \\
n \to p + e^{-} + \bar\nu_{\rm e} \nonumber \\
n + \nu_{\rm e} \to p + e^{-},
\end{eqnarray}
and their reaction rates are given by appropriate integrals 
\citep{reddy, janiuk2007}.
In addition, the neutrinos are produced due to the
 electron-positron pair anihillation, bremsstrahlung, plasmon decay, in the following reactions: 
\begin{equation}
e^{-}+e^{+}\to \nu_{\rm i}+\bar\nu_{\rm i} \nonumber \\
\label{eq:annihil}
\end{equation}
\begin{equation}
n+n \to n+n+\nu_{\rm i}+\bar\nu_{\rm i} \nonumber \\
\label{eq:brems}
\end{equation}
\begin{equation}
\tilde \gamma \to \nu_{\rm e}+\bar\nu_{\rm e} .\nonumber \\
\label{eq:plasmon} 
\end{equation}
We calculate their rates numerically, with proper integrals over the distribution 
function of relativistic, partially degenerate species.
Finally, the total neutrino cooling rate is given by the two-stream approximation \citep{dimatteo2002}, and it includes
the scattering and absorptive optical depths for neutrinos of the three
flavors: 
\begin{equation}
Q^{-}_{\nu} = { {7 \over 8} \sigma T^{4} \over 
{3 \over 4}} \sum_{i=e,\mu,\tau} { 1 \over {\tau_{\rm a, \nu_{i}} + \tau_{\rm s} \over 2} 
+ {1 \over \sqrt 3} + 
{1 \over 3\tau_{\rm a, \nu_{i}}}} \times {1 \over H}\;,
\label{eq:qnuthick}
\end{equation}
and is expressed in the units of erg s$^{-1}$cm$^{-3}$.

The structure of a magnetized accreting torus in which the above 
reactions take place and the gas
looses energy via neutrino cooling, was investigated in our previous paper 
\citep{janiuk13}. 
In that work, however, the neutrino cooling rate 
was used to update in every time step only the internal energy in the plasma, and the
pressure was still approximately computed 
from the adiabatic relation, $p=(\gamma-1)u$, with constant $\gamma=4/3$, which 
simplifies a lot the GR MHD scheme.

The leptons and baryons are relativistic in the GRB engine, 
and may have an arbitrary 
degeneracy level, so we compute the gas pressure using the
Fermi-Dirac distribution of the species. Also, neutrinos may contribute to the pressure
due to the fact that they are scattered and absorbed, similarly to the photons
and (rather negligibly small in the GRB case) radiation pressure component.
In the total pressure, we include now
the contributions from the free nucleons, pairs, radiation, trapped neutrinos, 
and also from the Helium nuclei.  
\begin{equation}
P_{\rm gas} = P_{\rm nucl}+P_{\rm He}+P_{\rm rad}+P_{\nu}
\end{equation} 
where $P_{\rm nucl}$ includes free neutrons, protons, 
and the electron-positrons:
\begin{equation}
P_{\rm nucl}=P_{\rm e-}+P_{\rm e+}+P_{\rm n}+P_{\rm p}
\end{equation}
with
\begin{equation}
P_{\rm i} = {2 \sqrt{2}\over 3\pi^{2}}
{(m_{i}c^{2})^{4} \over (\hbar c)^{3}}\beta_{i}^{5/2}
\left[F_{3/2}(\eta_{\rm i},\beta_{\rm i})+{1\over 2} \beta_{\rm i}F_{5/2}(\eta_{\rm i},\beta_{\rm i})\right].
\label{eq:pi}
\end{equation}
Here, $F_{\rm k}$ are the Fermi-Dirac integrals of the order $k$, and
$\eta_{\rm e}$, $\eta_{\rm p}$ and $\eta_{\rm n}$ are the reduced chemical
potentials, $\eta_i = \mu_i/kT$  
is the degeneracy
parameter, where $\mu_i$ is the standard chemical
potential. Reduced chemical potential of positrons is
$\eta_{\rm e+}=-\eta_{\rm e}-2/\beta_{\rm e}$. Relativity parameters are defined as $\beta_{\rm i}=kT/m_{\rm i}c^{2}$.
The total pressure is computed numerically by solving the balance of nuclear reactions \citep{yuan2005, janiuk2010}.

\subsection{Initial conditions and dynamical model}

The numerically computed nuclear EOS is incorporated now into the MHD scheme,
with the pressure determined
as a function of density and temperature. 
The MHD scheme  solves for the
inversion between the so called 'primitive' and 'conserved' 
variables at every time-step (see e.g., \citealt{noble}).
The code HARM-2D provides solver for continuity and energy-momentum conservation equations:
\begin{equation}
(\rho u^\mu)_{;\mu} = 0 
\end{equation}
and
\begin{equation}
 {T^\mu}_{\nu;\mu} = 0,
\end{equation}
where:
\begin{eqnarray}
\label{eq:tmunu}
 T^{\mu\nu} = T^{\mu\nu}_{\rm gas} + T^{\mu\nu}_{\rm EM}  \nonumber \\
 T^{\mu\nu}_{\rm gas} = \rho h u^\mu u^\nu + pg^{\mu\nu} =(\rho + u + p) u^\mu u^\nu + pg^{\mu\nu}  \nonumber \\
 T^{\mu\nu}_{\rm EM} = b^2 u^\mu u^\nu + \frac{1}{2}b^2 g^{\mu\nu} - b^\mu b^\nu ; ~~ b^\mu = u_{\nu}\dF^{\mu\nu} \nonumber \\
\end{eqnarray}
and $u^\mu$ is the four-velocity of gas, $u$ is the internal energy density,  
$b^\mu = \frac{1}{2} \epsilon^{\mu\nu\rho\sigma}u_\nu F_{\rho\sigma}$ is the magnetic four-vector, 
and $F$ is the electromagnetic stress tensor.
Assuming the force-free approximation, we have $E_{\nu}=u_{\mu}F^{\mu\nu}=0$.

The conservative numerical scheme which is used in this code, in general solves
$\del_t \bU(\bP) = -\del_i \bF^i(\bP) + \mathbf{S}(\bP)$
where $\bU$ is a  vector of ``conserved'' variables (momentum, energy density,
number density, taken in the coordinate frame), $\bP$ is a vector of
'primitive' variables (rest mass density, internal energy), 
 $\bF^i$ are the fluxes, and $\mathbf{S}$ is a vector of source terms. 

In contrast to the non-relativistic MHD, where $\bP \rightarrow \bU$ and  $\bU
\rightarrow \bP$ have a closed-form solution, in relativistic MHD,
$\bU(\bP)$ is a complicated, nonlinear relation. Inversion
$\bP(\bU)$ is calculated therefore, in every time step, numerically.
The inverse transformation requires to solve a set of 5 non-linear equations,
 which is done by a multi-dimensional 
Newton-Raphson routine.
The procedure is simple, if the pressure is given by an adiabatic relation with density. However, for a general equation of state, one must compute also $dp/dw$, $dp/dv^{2}$, and $p(W, v^{2},D)$ (here we have enthalpy, $w=p+u+\rho$; $W=\gamma^2 w$; $D=\sqrt{-g}\rho u^t$).
In our scheme, we have tabulated pressure and internal energy as a function of density and temperature, $(p,u)(\rho,T)$, and we interpolate over this table.
In addition, density and temperature in the flow give us a unique determination of the neutrino cooling rate.
The Jacobian $\partial \bU /\partial \bP$ is computed numerically, and  the conserved variables are evolved in time.
To speed up the computations, our version of this code was 
parallelized using the MPI
technique, for the hydro-evolution, and we also implemented the 
shared memory hyper-threading for the EOS-table interpolation.

The simulations are conducted in 2D, on the spherical grid 
with the resolution of  256x256 cells in the $r$ and $\theta$ directions, 
and typically run up to t=2000-3000 M in dimensionless units (c=G=1). 
The grid is spaced logarithmically in radius and concentrated towards the 
equatorial plane, defined as in \citet{gammie}.

The initial conditions for the accretion flow are given by the 
equilibrium torus solution, defined as in \citet{Moncrief} and \citet{Abramowicz}.
 The parameters of the model
are the black hole mass and its spin.
The torus mass is defined by the radius of the pressure maximum, and it 
is also
supplemented with the density scaling, which is necessary to
compute the pressure and temperature in the torus in physical units.
The model
is seeded by the initial magnetic field.
The magnetic field in our current computations 
 is adopted to have a standard, poloidal configuration given with the 
$\phi$-component of its vector potential scaling with the density, 
$A_{\phi}= (\rho/\rho_{max})$, and is weakly magnetized, with an
initial $\beta=P_{\rm gas}/P_{\rm mag}=50$ (see e.g., \citet{McKinney2012}
for the discussion of various other field configurations).

\section{Structure of the flow}\label{sec:results}

\subsection{Accreting torus and wind outflow}

\begin{figure}
\includegraphics[width=7cm]{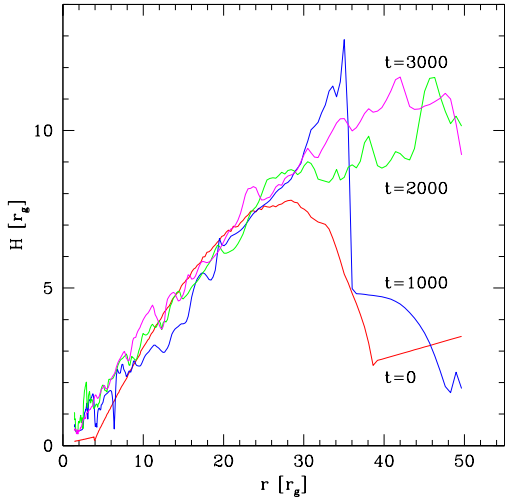}
\caption{Thickness of the torus in the function of radius, at four different time 
snapshots during the simulation, t=0, 1000, 2000, and 3000 M, as labelled in the plot and marked by different colors. 
The model includes numerical EOS and neutrino cooling. Parameters of the model are $M_{\rm BH}=3 M_{\odot}$, $M_{\rm torus}\approx 0.1 M_{\odot}$.
Black hole spin is  $a=0.9$.}
\label{fig:hdisk}
\end{figure}

Our choice of the setup and parameters is meant to reproduce the
properties of the typical engine of a short gamma ray burts, however, to some extent, the computations are scalable with the black hole and accretion disk mass.
Nevertheless, in the discussed models, we assume  no matter supply through the outer boundary to the computational domain, which would result, e.g., from a fallback of the collapsar's envelope in case of long GRBs. 
We choose in this Section a fiducial value of $M_{\rm BH}=3 M_{\odot}$, 
as an order of magnitude adequate to be a result of a compact binary merger.
The torus mass is taken here to be of about 0.1 $M_{\odot}$, so that the accretion rate in physical units is on the order of 1 Solar mass per second \citep{Kluzniak98}.  
The initial conditions introduced from the pressure equilibrium solution are relaxed
after about $1000 M$ (we express the time in dimensionless units, with $G=c=1$, so that the physical unit of time in seconds will be $t_{\rm unit}=GM/c^{3}$), and then
the torus evolves to the form the dynamically driven, geometrically
 thin flow, which is accreting
through the black hole horizon.
In the example Figure \ref{fig:hdisk},
we plot the radial profiles of the 
torus thickness for the cooled torus, at four time snapshots. The initial condition represents the equilibrium torus (a 'doughnut') and 
this profile spreads with time into larger radii. 
The torus thickness is computed from the pressure scale-height, as follows
\begin{equation}
H = {1 \over \Omega} c  \sqrt{P \over \rho c^{2}+e}, 
\label{eq:hdisk}
\end{equation}
where $\Omega = {c^{3} \over GM} {1 \over (a+r^{3/2})}$ is the Keplerian frequency and the sound speed is approximately given by its relativistic form
(see \citet{ibanez}).
The thickness of the torus is about $H/r\sim 0.2$. 
The exact value 
depends on the black hole spin, both in the initial equilibrium 
and time-dependent solution. The maximum thickness of the torus, defined above, reaches the values from 7 to 9 $r_{\rm g}$ in the initial model, for the 
spins from 0.6, to 0.98. The maximum thickness at the time $t=2000 M$ increases
to about 10-14 $r_{\rm g}$, for this range of spins. Also, as the Figure \ref{fig:hdisk} shows, it is shifted to the outer boundary, while the torus no longer keeps its equilibrium 'doughnut' shape.

\begin{figure}
\includegraphics[width=7cm]{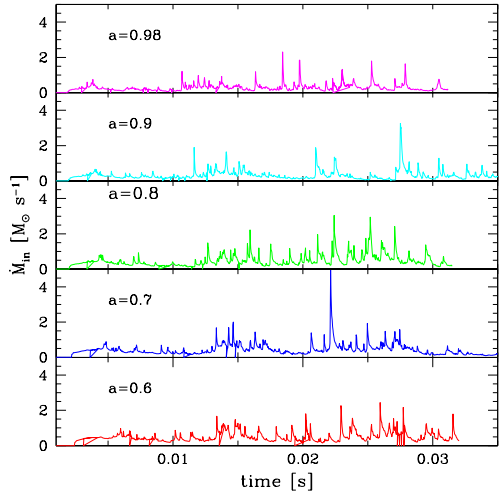}
\caption{Accretion rate as a function of time, in the models with
 neutrino
  cooling. The models were calculated for a range 
of black hole spin parameters: a=0.6, 0.7, 0.8, 0.9, and 0.98,
with red, blue, green, yellow, and magenta lines, respectively.  
  The black hole mass is equal to  $3 M_{\odot}$, and the
  initial disk mass is about  $0.1 M_{\odot}$}
\label{fig:accrate}
\end{figure}

In Figure \ref{fig:accrate} we plot the accretion rate through the
black hole horizon in the function of time. The mean rate of accretion 
is about 0.3-0.4 $M_{\odot}$ s$^{-1}$, for the black hole mass of 3 
$M_{\odot}$, depending slightly on the black hole spin (see Table~\ref{table:models}).
The instantaneous peaks which appear in the states of a highly variable accretion rate, reach the values
up to several Solar mass per second. These flares disappear as the initial 
condition is evolved, and then the accretion rate varies only by a few per cent.
The flares in the local accretion rate do not necessarily correspond to the observable luminosity flares. In fact, the neutrino luminosity which we derive by integrating the emissivity over the whole simulation volume, has a smooth dependence on time. In case if the high accretion rate through the black hole 
horizon affects somehow the jet production, this rapid 
variability might have observable consequences. The direct predictions however are not possible with the current model.

\begin{figure*}
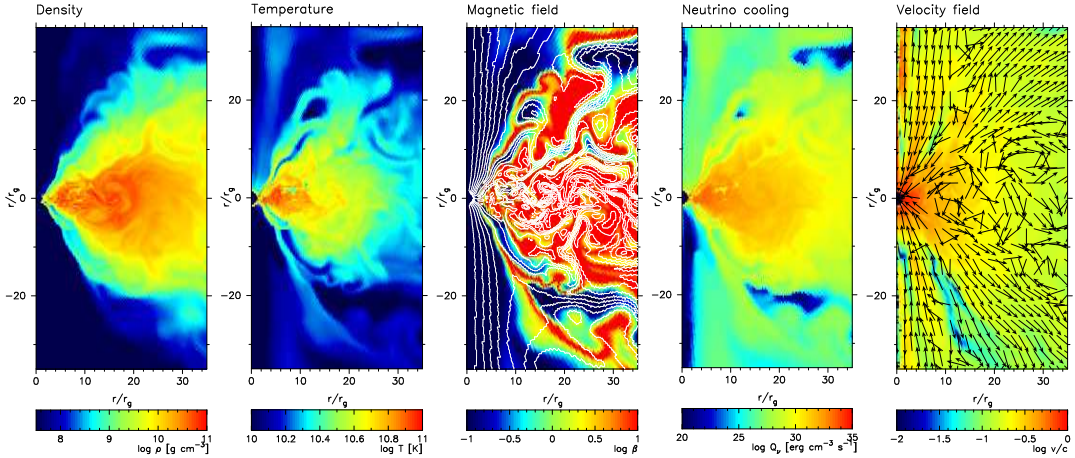

\centering
\includegraphics[width=6cm,angle=270]{map.rho.sbhAc.d200.ps}
\includegraphics[width=6cm,angle=270]{map.temp.sbhAc.d200.ps}
\includegraphics[width=6cm,angle=270]{map.beta.sbhAc.d200.ps}
\includegraphics[width=6cm,angle=270]{map.neu.sbhAc.d200.ps}
\includegraphics[width=6cm,angle=270]{map.vel.sbhAc.d200.ps}
\caption{Structure of accretion disk at the end of the simulation. 
The model accounts for the neutrino cooling
  and nuclear EOS.  
The maps show: (i) density,
  (ii) temperature of the plasma, (iii) ratio of gas to magnetic pressure,
  with field lines topology, (iv) neutrino emissivity $Q_{\nu}$, 
and (v) velocity field (from left to right).  
The snapshots are taken at time t=2000 M, which is equivalent to $t\approx 0.03$ s.
Parameters of the model: black hole mass $M_{\rm BH} = 3 M_{\odot}$, torus mass 
$M_{\rm t} = 0.1 M_{\odot}$, and BH spin  $a=0.6$.
\label{fig:mass3}}
\end{figure*}

In Figure~\ref{fig:mass3} we show the two-dimensional maps of the
torus structure, in the $r-\theta$ plane, 
obtained from the calculations with 
the black hole mass of 3 $M_{\odot}$ and spin of $a=0.6$.
The snapshots from our simulation were
 taken at time $t=2000 M$, which for the adopted black hole mass
 is equivalent to about 0.03 s.
The maps present the rest mass density $\rho$, gas
temperature $T$, and magnetic pressure to gas pressure ratio $\beta$, over-plotted with magnetic field
lines. We also show the distribution of the neutrino emissivities within the flow, 
as computed from the equation of state of matter with these profiles of 
 temperature and density. 
The last panel to the right in Figure \ref{fig:mass3}, presents 
the velocity field in the flow. The arrows (with normalized length) 
show the direction of velocity, while the color map presents the ratio of 
the flow velocity to the speed of light, $v/c=\sqrt{g^{\mu\nu}u^{\mu}u^{\nu}}$. 
The turbulent velocity field in the torus is changing into a uniform outflow 
which is launched above the torus surface at high latitudes.

The disk winds are sweeping the gas out from the system, both in the
equatorial plane and at higher latitudes. 
We identified the regions of the wind in the computation domain 
by defining three conditions that must be satisfied simultaneously. In order to
distinguish the wind from the torus main body, and from the magnetized jets: 
(i) the radial velocity of the plasma is positive 
(ii) the density is smaller than $10^{9}$ g cm$^{3}$
and (iii) the gas pressure is dominant, $\beta>1.0$. 
 The winds are located approximately at radii
above 10 $R_{\rm g}$ and 
latitudes between about $30^{\circ}-60^{\circ}$ and  $120^{\circ}-150^{\circ}$. 
 The velocity in the wind is typically $v/c \sim 0.06-0.5$, 
and its maximum value is only
slightly depending on the black hole spin. 
The density in the wind, by definition, is below $10^{9}$ g sm$^{-3}$, but
 may drop down to $10^{5}$ g cm$^{-3}$. The temperature
of the wind in the neutrino cooled models is in the range of 
 $6\times 10^{9} - 5.7\times 10^{10}$ K.
Such high temperatures, above the threshold for electron-positron pair production, $T=m_{\rm e}c^{2}\approx 5 \times 10^{9}$ K, are 
the key condition for neutrino emission processes. The neutrino cooling 
is then efficient, as it only weakly depends on density. In the clumps with 
$\rho > \sim 10^{8}$ g cm$^{-3}$, the nuclear processes lead to neutrino 
production, while the optical depths for their absorption are very small.

The effect of the wind is the mass loss from the system.  
We estimated quantitatively the
 evolution of the mass during
the simulation. 
The total mass removed from the torus during the simulation,
calculated by integrating the density over the total volume, 
differs from 
the total mass accreted onto the black hole
 (i.e. the time integrated mass accretion rate through the inner boundary,
subtracted from the initial mass).
This may depend on the black hole spin.
For the model with $a=0.6$, the simulation was ended when the total mass
in the volume dropped below $0.089 M_{\odot}$. The starting mass at $t=0$, was
equal to $0.11 M_{\odot}$. During the simulation, about 0.014 Solar masses was accreted onto the black hole (which makes 20\% of the initial mass), and the rest, about 6\%, was lost from the computation domain in the outflows.
For larger black hole spins, the fraction of mass lost through wind can reach about 10\%.
The values for all the models are collected in Table \ref{table:models}.

The hot, magnetized, transient polar
jets appear as well, and they form on both
sides of the black hole, as can be seen in 
the maps in Figs. \ref{fig:mass3}. They are present in a narrow 
region along the polar axis, where the open magnetic field lines have formed.
Velocity field is directed there outwards, and the gas is expelled from the black hole
region with the velocity that is nearly at the speed of light.

\subsection{Neutrino luminosity and Blandford-Znajek energy extraction}

\begin{figure}
\includegraphics[width=7cm]{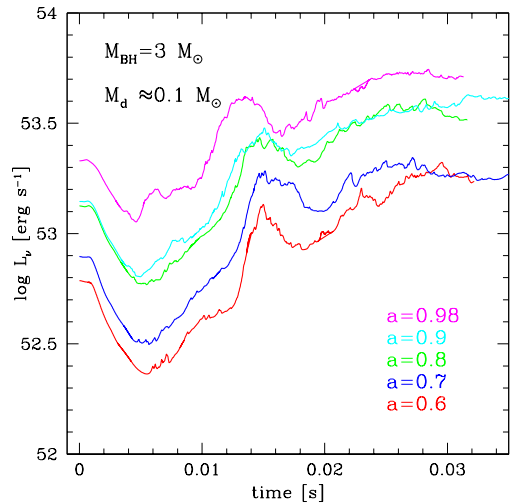}
\caption{Total neutrino luminosity as a function of time.  
The black hole mass is
  $M_{\rm BH}=3 M_{\odot}$ and torus initial mass is 
$M_{\rm d} \sim 0.1 M_{\odot}$.
The models were calculated for a range of 
black hole spin parameters, $a=0.98, 0.9, 0.8, 0.7$, and 0.6, as marked with
magenta, yellow, green, blue, and red lines, respectively.}
\label{fig:neurate3}
\end{figure}

The total neutrino luminosity in the function of time is shown in Figure
\ref{fig:neurate3}, where we present the results
for the black hole mass of 3 $M_{\odot}$
and torus mass of  $\approx 0.1 M_{\odot}$. The luminosities 
are plotted for a range of black hole spins: 
$a=0.6, 0.7, 0.8, 0.9$, and 0.98.
Initially, the total integrated 
neutrino luminosity 
is in the range of $6 \times 10^{52}$-$2 \times 10^{53}$ erg s$^{-1}$, 
and temporarily drops, when the matter starts accreting.  
As the initial state is evolved, the luminosity
grows to over $10^{53}$ erg s$^{-1}$ and at time $t=0.03$ s
reaches roughly a constant level, which is scaling with
the black hole spin.  
The exact values of $L_{\nu}$ at the
end of the simulation are given in Table~\ref{table:models}. 

The neutrinos are emitted from the torus as well as from the hot, rarefied
wind. The wind contribution to the total
luminosity is of about 10-11\% for all the
models.
The luminosity of the densest parts of the torus, estimated 
by weighing the total emissivity by the plasma density, is only
on the order of $10^{48}-10^{49}$ erg s$^{-1}$. This is because the total 
opacity for
neutrino absorption and scattering in this regions reaches $\tau \sim 1 - 5$,
and the neutrinos are trapped in the dense plasma.

The power and luminosity available through the
Blandford-Znajek process \citep{znajek1977}
is calculated  from the
electromagnetic stress tensor, given by
\begin{equation}
T^{\mu \nu}_{\rm EM} = b^{2}u^{\mu} u^{\nu} + \frac{b^{2}}{2} g^{\mu \nu} - b^{\mu} b^{\nu},
\end{equation}
where $b^{\mu}$ is the magnetic four-vector, with $b^{t}=g_{i \mu}B^{i}u^{\mu}$ and $b^{i}=(B^{i}+u^{i}b^{t})/u^{t}$ and $u^{\mu}$ is the four-velocity. We compute the radial electromagnetic flux through the horizon as in
\citet{McKinneyGammie2004}:
\begin{equation}
\dot E = 2\pi \int_{0}^{\pi} d \theta \sqrt{-g} F_{\rm EM},
\end{equation}
where $F_{\rm EM} = - T^{r}_{t}$ and $g$ is the metric determinant.
We consider only the flux, which is dragged inwards through the black hole horizon, and we also check the magnetization of the plasma there,
$\beta(r_{\rm in})=B^{2}/\rho(r_{\rm in})$, where $\rho$ is the rest mass
density of the gas. We impose a condition of $\beta(r_{\rm in}) \ge 1.0$, otherwise the Blandford-Znajek power is set to be zero.
This is because we want to focus on 
the energy extracted to the magnetically dominated polar jets. The outward directed energy flux can be in fact present
also close to the equator, at moderate altitudes, where the matter is dense 
and weakly magnetized. The contribution of this equatorial 
flux to the total power is only about 5-15\%, if the black hole spin is $a \ge 0.9$, however it can be dominant for the models with small black hole spins, 
when the magnetized jets are not formed. 
In some runs, the dense and transient clumps of matter temporarily appear in the polar regions, so that 
the jet is contaminated and its BZ luminosity temporarily vanishes 
(see Figure 5). Our condition is therefore numerically equivalent to the 
method of jet 'cleaning', used by other authors (see e.g. \citealt{TchBrom2016}).

\begin{figure}
\includegraphics[width=7cm]{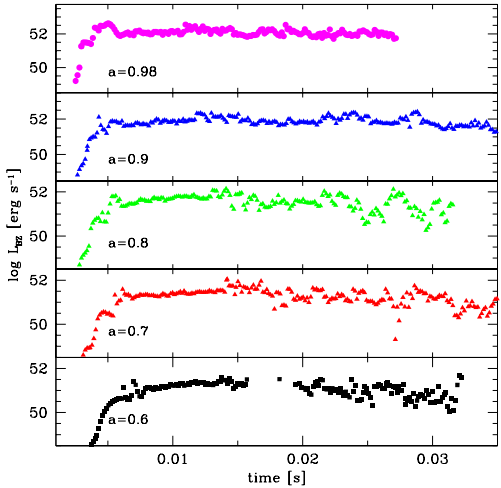}
\caption{Power extracted from the black hole rotation through Blandford-Znajek process, and resulting luminosity as a function of time. The model includes numerical EOS and neutrino cooling. Parameters of the model are $M_{\rm BH}=3 M_{\odot}$, torus mass $M_{\rm t}\approx 0.1 M_{\odot}$. 
The black hole spin values are marked in the panels, and range from a=0.6, to a= 0.98, from bottom to top panels}
\label{fig:lbz_sbhc}
\end{figure}

The resulting BZ luminosities in the function of time for the range of black hole spin values in different models are presented in Figure \ref{fig:lbz_sbhc}.
As the Figure shows, the continuous power supply to the jet of above $10^{52}$ erg s$^{-1}$ is available only for the highly spinning black holes.
If the spin is moderate, $a=0.6-0.8$, then the power $L_{\rm BZ}\approx 10^{51}$ erg s$^{-1}$ is injected. Episodically, the power drops to zero, when there is no magnetization on the black hole horizon. Therefore, the power injection can be separated by the intervals of the 'BZ-quiescence', which can be longer if the black hole spin is small.

The exact values of the BZ-luminosity at the end of the simulations are given in Table \ref{table:models}.

\small{
\begin{table*}
\begin{center}
\caption[]{Summary of the models. Black hole mass is
  $3 M_{\odot}$. The mass of the torus is given in $M_{\odot}$, accretion rate in 
$M_{\odot}$s$^{-1}$, the time is in seconds, and luminosity in erg s$^{-1}$. \label{table:models}
}
\begin{tabular}{ccccccccccc}
\hline
Model & $a$ & $R_{\rm max}$ & $t_{\rm e}$ & $M_{\rm t}(t_{0})$ &  $M_{\rm t}(t_{\rm e})$ & $<\dot M>$ & $\dot M(t_{\rm e})$ & $\Delta M_{\rm w}$ & $L^{\rm tot}_{\nu}(t_{\rm e})$ & $L_{\rm BZ}(t_{\rm e})$ \\ 
\hline   
sAc & 0.6 & 11.8 & 0.028 & 0.11 & 0.082 & 0.49 & 0.30  & 0.012 & $ 1.69 \times 10^{53}$ & $1.83 \times 10^{50}$ \\
sBc & 0.7 & 11.0 & 0.029 & 0.10 & 0.072 & 0.45 & 0.29  & 0.014 & $ 1.67 \times 10^{53}$ & $2.99 \times 10^{51}$ \\
sCc & 0.8 & 10.5 & 0.029 & 0.12 & 0.086 & 0.46 & 0.35  & 0.018 & $ 3.40 \times 10^{53}$ & $ 5.45 \times 10^{50}$ \\
sDc & 0.9 & 9.75 & 0.028  & 0.10 & 0.075 & 0.31 & 0.22  & 0.015 & $ 3.62 \times 10^{53}$ & $ 2.59 \times 10^{52}$ \\
sEc & 0.98 & 9.1 &  0.028 & 0.11 & 0.083  & 0.28 & 0.20  & 0.017 & $ 5.25 \times 10^{53}$ & $ 6.6 \times 10^{51}$ \\
\hline
sAnc & 0.6 & 11.8 & 0.032 & 0.11  & 0.090 & 0.46  & 0.31  &  0.005 & -- & $ 1.0 \times 10^{49}$ \\
sBnc & 0.7 & 11.0 & 0.031 & 0.10  & 0.084 & 0.37  &  0.18  & 0.004 & -- & $ 2.8 \times 10^{50}$ \\
sCnc & 0.8 & 10.5 & 0.044  & 0.12 & 0.108 & 0.24  &  0.04 & 0.004 & -- & $ 4.43 \times 10^{50}$ \\
sDnc & 0.9 & 9.75 & 0.041 & 0.10 & 0.089 & 0.23  & 0.177   & 0.003 & -- &  $ 2.52 \times 10^{51}$\\
sEnc & 0.98 & 9.1 & 0.035 & 0.11 & 0.097 & 0.18 & 0.109  &  0.007 & -- & $ 7.62 \times 10^{51}$ \\
\hline
\end{tabular}
\end{center} 
\end{table*}
}

\subsection{Comparison to the models without neutrino cooling}

To quantify the effect of microphysics, equation of state, and the neutrino
cooling in 2D MHD simulations, we ran a set of test models with no cooling 
and adiabatic equation of state with $\gamma=4/3$, as in the standard
HARM models. The model parameters and initial conditions were otherwise 
the same.

\begin{figure}
\includegraphics[width=7cm]{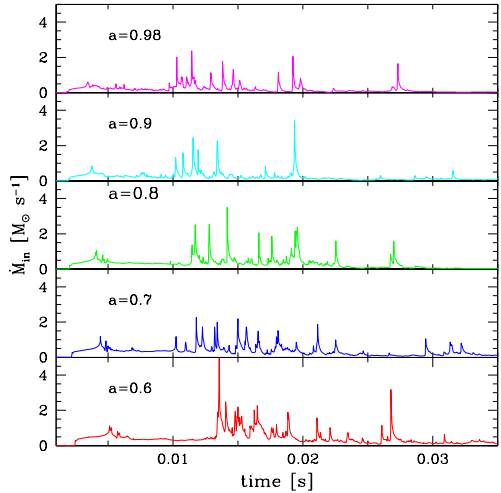}
\caption{Accretion rate as a function of time, in the models 
without neutrino cooling, with polytropic EOS adopted with $\gamma=4/3$.
The values of black hole mass, spin, and disk mass are denoted in the figure.}
\label{fig:accrate_nc}
\end{figure}

In Figure \ref{fig:accrate_nc} 
we plot the rate of matter accretion through the inner boundary, 
 as a function of
time for a set of models with various black hole spins.  
The average accretion rate onto
black hole is slightly lower in the adiabatic models than in the cooled models, 
for the same
model parameters.  The initial conditions are evolved slower in the adiabatic 
case, and the flaring peaks in the accretion rate appear for a longer time. 
The highest peak in this set of models appeared for $a=0.6$, and not for $a=0.7$, which was the case in the neutrino cooled torus.
The quasi-stationary, almost constant accretion rate
appears after $t\approx 0.04$, and the mean value of the accretion rate
 is on the same order as in neutrino cooled models, but somewhat lower (see Table \ref{table:models} for exact values).

\begin{figure}
\includegraphics[width=7cm]{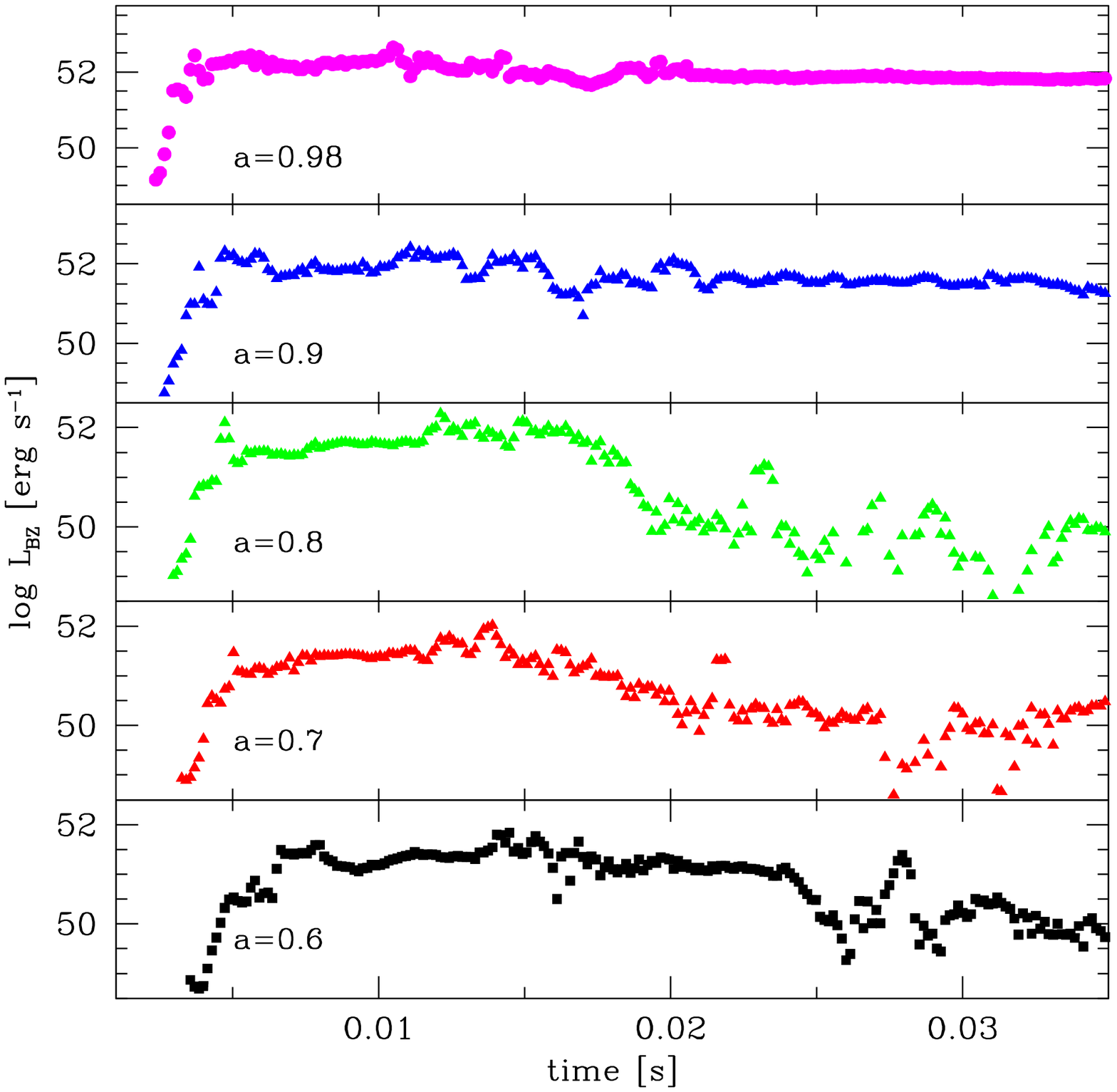}
\caption{Power extracted from the black hole rotation through Blandford-Znajek process, and resulting luminosity as a function of time. The model is based o the adiabatic EOS, with $\gamma=4/3$, and no neutrino cooling. Parameters of the model are $M_{\rm BH}=3 M_{\odot}$, torus mass $M_{\rm t}\approx 0.1 M_{\odot}$. 
The black hole spin values are marked in the panels, and range from a=0.6, to a= 0.98, from bottom to top panels}
\label{fig:lbz_sbhnc}
\end{figure}

In the adiabatic models, the neutrino luminosity is set to zero, by definition.
The only source of the energy extraction and power supply to the jets is therefore the rotation of the black hole and magnetic flux.
In Figure \ref{fig:lbz_sbhnc} we show the Blandford-Znajek luminosity in the function of time for this set of models, with various spins.
We note that for the highest black hole spins, $a=0.9-0.98$, the BZ luminosity is equally large as for the neutrino cooled tori, and it is 
almost constant with time. However, for the models with black hole spins of $a=0.6-0.8$, the BZ power drops later in the evolution, and after $t\approx 0.02$ it is smaller than about $10^{50}$ erg s$^{-1}$. In other words, the energy extraction from the moderately rotating black hole by magnetic fields is less efficient, when the torus is not cooled by neutrinos. 

\begin{figure*}
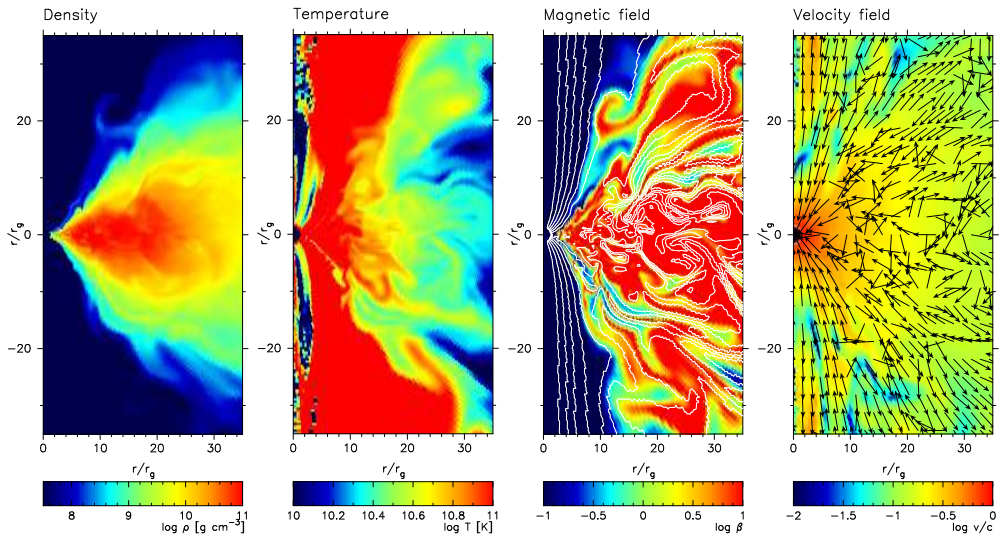

\centering
\includegraphics[width=7cm,angle=270]{map.rho.sbhAnc.d200.ps}
\includegraphics[width=7cm,angle=270]{map.temp.sbhAnc.d200.ps}
\includegraphics[width=7cm,angle=270]{map.beta.sbhAnc.d200.ps}
\includegraphics[width=7cm,angle=270]{map.vel.sbhAnc.d200.ps}
\caption{Model without neutrino cooling and with adiabatic equation of state. The parameters are: $a=0.6$,
  $M_{\rm BH} = 3 M_{\odot}$, $\beta_{\rm init}=50$, torus mass 
$M_{\rm t} \approx 0.1  M_{\odot}$.  
The maps, from left to right, show the distribution of density,
  temperature, ratio of gas to magnetic pressure with field lines topology,
  and velocity field. The snapshots are taken at $t=2000 M$, which is equivalent to $t\approx0.03$ s of the dynamical simulation.}
\label{fig:torus_nocool}
\end{figure*}

The maps of the density, temperature, velocity and
magnetic field in the exemplary model with $a=0.6$, are shown in Figure \ref{fig:torus_nocool}.
The density of the disk in the models without cooling is
larger in the equatorial plane, 
while the disk seems to be compact (i.e., denser
and geometrically thinner) close to the black hole. Due to the
lack of neutrino cooling, it is also hotter than that presented in the Figure 
\ref{fig:mass3}.  The highest temperatures are achieved in the regions below 10 $r_{\rm g}$ in the disk, and also in the outflows (winds).
Open magnetic field lines are formed, even for the small black hole spin, 
but in general the flow is less magnetized and the gas pressure is large both in the disk and in the winds. The velocity field map shows the cooler, low density outflowing clumps, confined along the vertical axis.

\begin{figure}
\includegraphics[width=7cm]{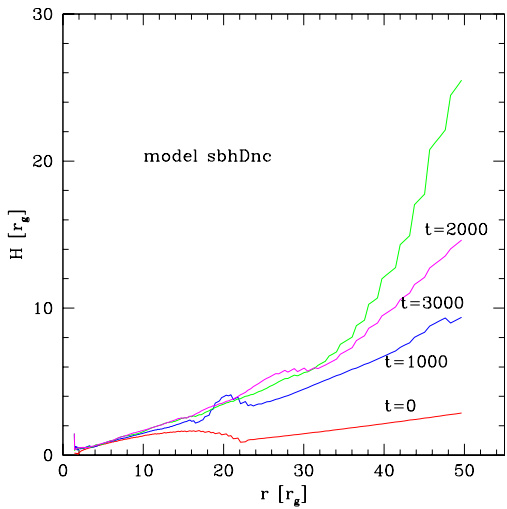}
\caption{Thickness of the torus as a function of radius, in the models 
without neutrino cooling, with polytropic EOS adopted with $\gamma=4/3$.
The model is with the black hole spin $a=0.9$.
}
\label{fig:hd_snc}
\end{figure}

The thickness of the torus, measured 
by the scale height given by Eq. \ref{eq:hdisk} is shown in the Figure \ref{fig:hd_snc}. 
In the adiabatic models, the disk is thicker geometrically, and at outer regions the ratio $H/r$ saturates at the value of about 0.3-0.4. 
The mass loss rate is smaller than in the case of neutrino cooled tori, and
typically only 1-3\% of the torus mass is lost in the wind 
during the simulation. For the model with black hole spin close to maximal, $a=0.98$, the total mass loss in the wind outflow is completely negligible, within our numerical accuracy.

\section{Nucleosynthesis of heavy elements}
\label{sec:nucleo}

In the astrophysical plasma, thermonuclear fusion occurs 
due to the capture and release of particles ($n$, $p$, $\alpha$, $\gamma$).
 Reaction sequence produces further isotopes, and
the set of non-linear differential equations is solved by the Euler method 
\citep{Meyer94,wallerstein1997}.
The nuclear reactions may proceed with 1 (decays, electron-positron capture, photodissociacion), 2 (encounters), or 3 (triple alpha reactions) nuclei.
Abundances of the isotopes are calculated under the assumption of nucleon number and charge conservation for a given density, temperature and electron fraction ($T \le 1 MeV$). Integrated cross-sections depending on temperature $kT$ are determined with the Maxwell-Boltzmann or Planck statistics, and the background screening and degeneracy of nucleons must be taken into account.

We use the thermonuclear reaction network code 
 \textit{http://webnucleo.org},
 and determine the relative abundances of heavy elements
in the accreting torus. Given the density, temperature, and electron fraction,
the {\it nuceq} library is used, and we compute the nuclear statistical 
equilibria
established for the thermonuclear fusion reactions, taking 
the reaction rates available on JINA {\it reaclib} online database
 \citep{HixMeyer06,Seiten08}. 
This network is appropriate for temperatures below $1 MeV$, which is the case at the outer radii of accretion disks in GRB engines.
The mass fraction of the isotopes is solved for converged profiles of density, 
temperature and electron fraction in the disk, defined as:
\begin{equation}
Y_{\rm e}=\frac{n_{\rm e^{-}}-n_{e^{+}}}{n_{\rm b}}.
\end{equation}
We use here a standard, publicly available software, and a
more detailed description of our procedure can be found in 
\citet{aj2014}.

\begin{figure}
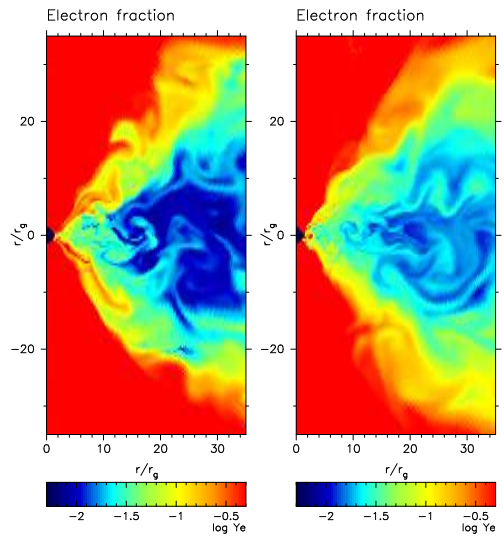

\includegraphics[width=7cm, angle=270]{map.ye.sbhAc.d200.ps}
\includegraphics[width=7cm, angle=270]{map.ye.sbhDc.d200.ps}
\caption{Electron fraction distribution. The snapshots are taken at $t=2000 M$. Models  shown in the left and right panels are $sAc$ and $sDc$, as labeled in the Table 1, respectively, and differ with the black hole spin.}
\label{fig:Ye2D}
\end{figure}

In Figure \ref{fig:Ye2D} we plot the electron fraction in the 2-D maps of the
accretion flow. 
The parameters of this simulation were: the black hole of mass equal to $3 M_{\odot}$, torus mass of about $0.11 M_{\odot}$, and the black hole spins of 0.6 and 0.9, as shown in the left and right panels, respectively. 
Distributions are plotted for the evolved simulation, 
at $t=2000 M$. 

\begin{figure}
\includegraphics[width=8cm]{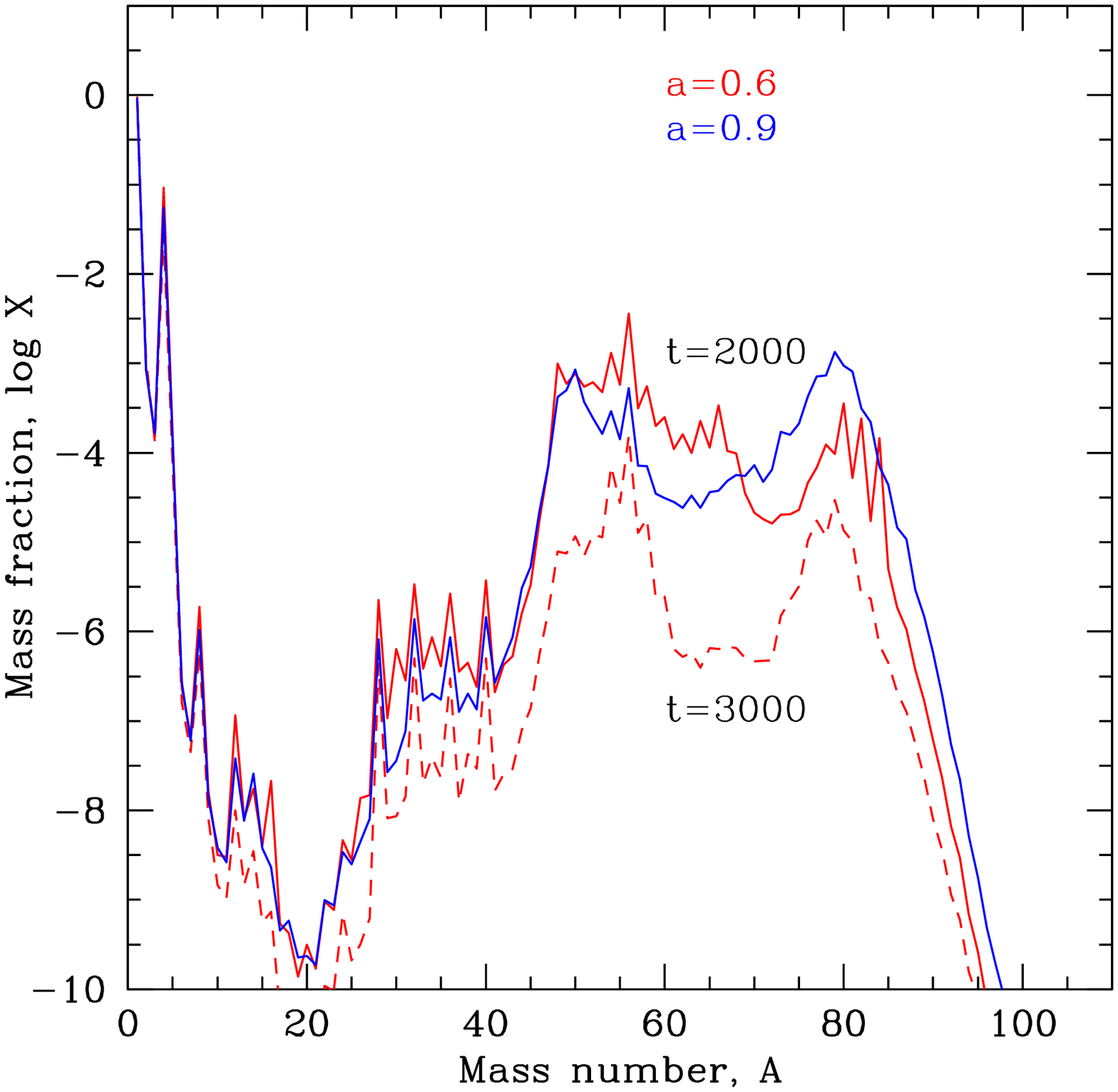}
\caption{
Volume integrated abundance distribution of elements 
synthesized in the 
accretion disk in 2D simulation with 
The black hole spin in the simulation was $a=0.9$ (blue line), and $a=0.6$ (red lines).
The distributions of density, temperature, and electron fraction in the flow 
were taken at two time snapshots: $t=2000 M$ and $t=3000 M$, as marked with solid and dashed 
lines, respectively.
}
\label{fig:nse2D}
\end{figure}

In Figure \ref{fig:nse2D} we show the abundances of heavy elements,
for these two models, which are
produced in the torus and wind. The numbers are
 integrated up to the distance of
$1000 R_{\rm g}$ from the black hole. 
The first setup corresponds to a weak GRB scenario, as the 
power extracted from the black hole rotation is in this case of only about $10^{50}$ erg s$^{-1}$, and the total energy emitted by neutrinos is of about $10^{53}$ erg s$^{-1}$. Taking into account the efficiency of conversion 
between neutrino-antineutrino annihilation process and the kinetic energy of the GRB jet, these two 
sources of power would result in a rather faint gamma ray burst.
In this GRB engine, along with the abundant light elements such as Carbon, 
and then Silicon, Sulfur and Calcium isotopes, we expect also to
find copious amounts of Titanium, Iron, and Nickel isotopes.
The amount of the heavy elements decreases with time, and
after about $0.02$ second, the abundances of these isotopes will drop by even 
10-100 times.

The second setup corresponds to a bright GRB scenario, when the power available from the Blandford-Znajek process is 100 times higher, than in the first case, 
and also the power from the neutrino annihillation is almost 3 times higher.
In this engine, similarly to the previous one, there will be produced isotopes of Iron and Vanadium, as well as Selenium, Manganese, Strontium, and Krypton.
The latter are about 10 times more abundant, than in the case of a faint GRB engine, and they can be produced efficiently in the engine for a longer time.

The isotopes synthetized in the GRB central engine accreting torus and its outflows during the prompt emission phase, should be detectable via the
X-ray emission that originates from their radioactive decay. 
These isotopes, such as $^{45}Ti$, $^{57}Co$, $^{58}Cu$, $^{62}Zn$, $^{65}Ga$, $^{60}Zn$, $^{49}Cr$, $^{65}Co$, $^{61}Co$, $^{61}Cu$, and $^{44}Ti$, 
might give the signal in the 12-80 keV energy band, which 
could in principle be observed by current instruments with a good energetic 
resolution, such as e.g., NuSTAR.

\section{Summary and discussion}\label{sec:diss}

We calculated the structure and short-term evolution of a gamma ray burst
central engine in the form of a magnetized torus accreting onto a black hole.
Our computations were performed in a fixed background metric around a 
spinning black hole, and the physical conditions in the plasma driven by the hyper-Eddington accretion rate were taken into account by the adopted 
microphysics. We describe the flow properties using the numerical equation of state of dense and hot matter, where the free nucleons and electron-positron pairs are relativistic and partially degenerate.
We estimate the resulting neutrino luminosity, which cools the flow via the nuclear reactions in which neutrinos of the three flavors are produced.
We conclude that the total integrated luminosity is larger than the power possibly extracted from the rotating black hole via the Blandford-Znajek process. 
Taking into account the uncertainty of the efficiency of the 
neutrino anihillation in the bi-polar jets \citep{zalamea2011}, we still find that 
the power supply to the jets by neutrinos is plausible and energetically competitive to the magnetic extraction of the black hole rotation energy.

The discussion of interplay between the Blandford-Znajek mechanism versus neutrino power supply to the GRB jets has been proposed in a number of works based on the so-called NDAF disk, e.g., \citet{liu2015}. These conclusions are however
far from robust, regarding the very simplified treatment of accretion flows in these models, which is described by a classical 1-dimensional steady state accretion flow with $\alpha$ viscosity, and pseudo-Newtonian gravity.
Our computations realistically treat the nuclear equation of state, together with other relevant physical ingredients, like the 
general relativity and magnetic fields. The latter is still lacking in some of the otherwise advanced, three-dimensional dynamical simulations of neutrino-cooled accretion flows
in the short GRB engines \citep{Dessart2009, Perego2014}. The essentially inviscid nature of such models might lead to their somewhat underestimated results for the accretion speed, dissipation and cooling rates.  

On the other hand, the relativistic models which are based on the ideal gas EOS (e.g., \citet{rezzola2011} use the adiabatic EOS with index of 2), can predict overestimated profiles of the disk scaleheight. Also, as noted by \citet{kyuchi2015}, who use the gamma index of 1.8 in their accretion flow computations, 
the value of this index will affect the amount of matter ejected in the torus wind. By accounting for the proper microphysics, instead of the simplistic EOS, we can get therefore most reliable description of the conditions in neutrino driven winds.
The exact determination of the accretion disk thickness is crucial with respect to the possible gravitational or thermal instabilities \citep{Perna2006, janiuk2007}, which 
can lead to the episodic accretion events and hence observable flares. As it was shown by \citet{BegelmanPringle2007},
the strong magnetic fields can be important for the stabilisation of the accretion flow against self-gravity. The ordered magnetic field configuration, which is mostly toroidal in the accretion torus with but predominantly poloidal and jet-like along the BH spin axis, is produced in simulation of \citet{rezzola2011} at time about 6 ms after the black hole formation. In our simulations the poloidal field lines appear after about 15 ms for the black hole of a 3 $M_{\odot}$, so the results are consistent within an order of magnitude. The difference in the timescale may be connected to some extent with the interplay between magnetic field dynamics and gas properties, but also affected by the 2-dimensional setup which is used in our model. We plan to expand the current computations into 3-dimensions in the future work, and then test quantitatively the resulting models against the observables of GRBs, such as their time variability, duration, and energetics. The ultimate impact of such simulations can also lead to the constraints on the magnetic fields strengths and configuration in the cosmic environment related to GRB progenitors.

Finally, based on the computed structure of the flow, its density, temperature, and electron fraction, we are able to
estimate the integrated abundances of heavy elements which are synthetized in the flow via the subsequent nuclear reaction chain.
These elements, if decaying radioactively at some distance from the central engine, will then contribute to the heating of the circum-burst medium, and afterglow emission.

As shown by \citet{sekiguchi2015}, the dynamical ejecta including neutrino transport may develop the shock waves, 
and related heating is responsible for enhanced reaction rates. The electron fraction is then increased, and their 
nuclear reaction networks are able to produce the r-process elements up to A=195. This was however not encountered 
in the Newtonian simulations, which produce elements up 
to A=130 \citep{goriely2011}.
Neutron star mergers are therefore expected to produce significant 
quantities of neutron-rich radioactive species, whose decay should result 
in a faint transient emission. The latter is called a `kilonova', and should appear in the days following the burst. 
The term refers to the luminosity peak in the near infrared band, rather than UV or Optical \citep{lipaczynski1998}, 
and the emission produced in the regions of the opacities much higher than in the case of of supernovae.
An observational support for such event has been reported e.g. by \citet{tanvir2013} for a short duration GRB 130603B.
Recently, re-examination of the observed long burst emission in GRB 060614 has shown that it is consistent with the 'kilonova' scenario, rather than an undelying supernova \citep{Yang2015}. 

The gamma ray bursts can play a unique role in the metal enrichment of the Universe, as they are the brightest sources at all redshifts, 
and, for long GRBs, they occur in the star forming regions.
Because the nucleosynthesis yields from the explosions of stars in different populations could differ by orders of magnitude, 
the metal abundance patterns, imprinted in the interstellar medium, should depend on the initial mass function
\citep{heger}.
The abundance patterns 
measured with distant GRBs can be used to determine the typical masses of the early stars (Pop III), and disentangle them from the younger Pop II SNe.
Such studies 
are planned e.g. with the newly developing instruments, e.g., on-board the Athena X-ray satellite, which is aiming to probe the GRB afterglow 
spectra, in combination with the studies of the quasars on the line of sight
\citep{jonker2013}.
The GRBs can be observed up to the highest redshifts (e.g., z=9.4, for GRB 090429, \citet{Cucchiara2011}) and hence the old stellar populations can be traced in their X-ray afterglow spectra, e.g., with absorption lines of Fe, Mg, Si, S, Ar, from the ionized gas in the GRB environment. 
Nevertheless, these data can be contaminated by the effects of the intergalatic medium and hence the observed absorption column is a combined effect of the warm-hot intergalactic medium, Lyman alpha clouds, as well as 
the circumburst medium intrinsic to the host galaxy \citep{Starling2013}. 

From the point of view of stellar evolution models, \citet{heger} computed the nucleosynthesis yields of elements produced in the massive metal-free stars, but they ignored both the neutrino-powered winds and the gamma-ray bursts accretion disks. The latter, as studied by \citet{Surman2006}, have large overproduction factors 
in their outflows for the nuclei like $^{44}Ti$, $^{45}Sc$, and $^{64}Zn$. Also, the light p-nuclei, such as $^{92}Mo$, and $^{94}Mo$, are produced in the rapidly accreting disks, as found by \citet{Fujimoto2003}).
The neutrino-driven winds, on the other hand, were modeled recently by \citet{Perego2014} in the context of binary neutron star mergers. They found that the
wind can successfully contribute to the weak r-process in the range of atomic masses from 70 to 110.

Obviously, the long GRBs are only a fraction of the massive star explosions and luminous supernovae \citep{Guetta2007, HjorthBloom2012}, and their impact for the majority of elements may not be crucial. However, verifying the signatures of these elements which are produced mainly in the GRB engines, may in the future occur to be a prospective tool to study the chemical evolution in the Universe due to the collapsing massive stars, via the available GRB afterglow X-ray spectra. 
Also, some of the bursts may be connected with peculiar type of events, such as the burst GRB 111209A/SN 2011kl
\citep{Kann2016}. In such case, the conditions in GRB central engine, determined during the explosion, can also lead to peculiar distributions of electron fraction, and affect the synthesis process for neutron-rich isotopes.

\begin{table*}
\begin{center}
\caption[]{Summary of the models with a massive black hole,
$M_{\rm BH}=62 M_{\odot}$, as estimated for the first LIGO-detected black hole. 
The mass of the torus is given in $M_{\odot}$, accretion rate in 
$M_{\odot}$s$^{-1}$, the time is in seconds, and luminosity in erg s$^{-1}$. 
\label{table:models_lbh}
}
\begin{tabular}{cccccccccc}
\hline
Model  & $a$ & $R_{\rm max}$ & $t_{\rm e}$ & $M_{\rm t}(t_{0})$ &  $M_{\rm t}(t_{\rm e})$ & $<\dot M>$ & $\dot M(t_{\rm e})$ & $L^{\rm tot}_{\nu}(t_{\rm e})$ & $L_{\rm BZ}(t_{\rm e})$ \\ 
\hline   
lbhAc & 0.6 & 9.1 & 0.91 & 16.8 & 14.36 & 5.01 & 4.06    & $4.80 \times 10^{53}$ & 0 \\
lbhBc & 0.7 & 8.4 & 0.91 & 13.0 & 11.17 & 4.64 & 1.99    & $6.18 \times 10^{53}$ & 0 \\
lbhCc & 0.8& 7.5 & 0.92 & 10.6 & 9.02 & 4.07 & 3.59    & $6.60 \times 10^{53}$ & $ 1.93 \times 10^{51}$ \\
lbhDc &  0.9& 7.0 & 0.91 & 10.4 & 8.89 & 3.98 & 2.55   & $9.72 \times 10^{53}$ & $ 3.30 \times 10^{51}$ \\
lbhEc &  0.98& 6.6 & 0.91 & 16.4 & 15.23 & 2.93 & 1.85   & $5.64 \times 10^{54}$ & $3.31 \times 10^{52}$ \\
\hline
\end{tabular}
\end{center} 
\end{table*}

The synthesized heavy elements can also be the source of a faint emission of a kind of an 'orphan' afterglow, if the main GRB jet is directed off-axis, or 
it is intrinsically very weak in its gamma ray fluence.
Still, the detection of such an afterglow emission will be a hint of the existence of a black hole in the engine site. Measurements of nucleosynthesis signatures may therefore help constrain the model parameters for such engine, namely the black hole spin and its mass.

The recent detections of the gravitational wave signal \citep{Abbott2016} 
confirmed independently the existence
of black holes in the Universe. The analysis of the gravitational wave signal allows to constrain the black hole mass and spin (with some uncertainty).
In the case of the source GW150914, it was found that the signal is related to the merger of a binary black hole. 
So far, the electromagnetic counterpart for this first event 
was tentatively reported only by Fermi satellite, who gave a limit for 
detection of a 
weak gamma ray signal \citep{Fermi2016}. No infrared emission was reported, and also
the signal was not confirmed by other gamma-ray missions for the case
of this burst \citep{greiner}.

Nevertheless, the detection of the electromagnetic counterpart of a gravitational wave source in the future events is appealing.
The related estimates for the heavy
elements synthesized in the ejecta could help constrain the parameters
for the GRB engine, if the latter happens to coincide with the gravitational wave source. 
In particular, if the binary black holes merge within a remnant circumbinary disk, which is a 
leftover after the past supernova explosion 
\citep{perna2016}, then the nucleosynthesis of elements within the disk engine and 
ejected winds can also be responsible for an infrared counterpart. 
Direct measurements of the escaping gamma-rays, which will be produced
at larger distance in the r-process nucleosynthesis, 
are probably difficult with current X-ray and gamma-ray missions \citep{Hotokezaka2016}.
 
The black hole, which was the product of the merger of two black holes of 
about 29 and 36 Solar mass black holes, was also a moderately spinning one.
The constraints for its dimensionless spin parameter obtained from the
amplitude and phase evolution of the observed gravitational waveform 
gave a value of $0.67^{+0.05}_{-0.07}$, while the final black hole mass was equal to
$62^{+4}_{-4}\ M_\odot$.
As can be shown, such a binary black hole merger, if accompanied by the matter accretion onto a 
rotating black hole, which happens either before, or after the merger, 
can lead to an unusual gamma ray burst \citep{janiuk2013b,janiuk2016}.
In the case of the relatively small rate of the black 
hole rotation, such as in GW150914, the accretion will result  
in a very small electromagnetic flux dragged through the
black hole horizon and hence a negligibly small power of the Blandford-Znajek process. 

We computed an additional set of the gamma ray burst engine models, with parameters
taken to be representative for the above scenario 
(see Table \ref{table:models_lbh}).
\begin{figure}
\includegraphics[width=8cm]{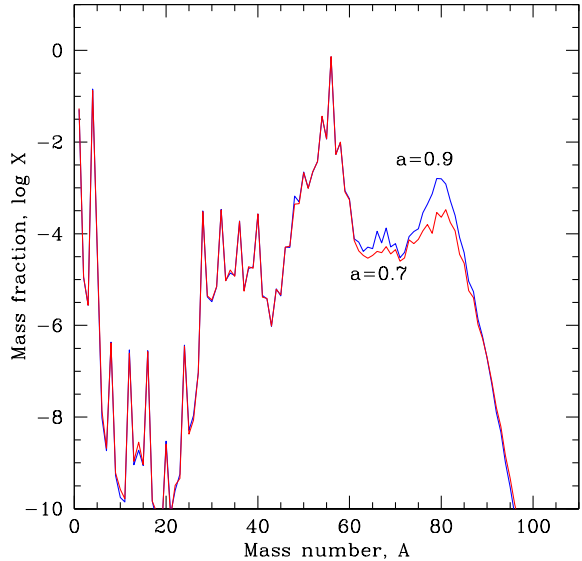}
\caption{Heavy elements synthesized in the torus and its wind, for the model lbhBc and lbhDc in the Table 2. The black hole mass is
 $M_{\rm BH} = 62 M_{\odot}$, and its spin is a=0.9 (blue), or a=0.7 (red).
The torus mass is about 15  $M_{\odot}$ and results were taken at time $t=2000 M$ in the dynamical simulation.}
\label{fig:nse2Dm62}
\end{figure}
We have also computed the mass fraction distribution of heavy elements produced in this putative engine (Figure \ref{fig:nse2Dm62}). We found, that contrary to the
presented in the previous Section 'fiducial' models, with a standard (small) black hole mass and 
accretion torus, here the nuclear conditions preferably
lead to production of mainly Iron peak elements. In the model \textit{lbhD} (see Table \ref{table:models_lbh}), 
the mass fraction of Iron 56 reached almost 1.0, while the isotopes from the group of Selenium and Manganese were $10^{3}$ times less abundant. 
This result suggests a possible additional test for the conditions present in the GRB engine, 
e.g., if the event happened to be coincident with a merger of the two 
$\sim 30 M_{\odot}$ black holes with masses of about $\sim 30 M_{\odot}$ each. It is also worth to notice that the supernova explosion which might have left a 30 Solar mass black hole, should have originated from at least 80 Solar mass star on zero-age main sequence, in a low-metallicity environment \citep{Spera2015}. Even if a significant amount of the star's mass was ejected during the evolution and explosion, some remnant disk is plausible, but detailed modeling of its structure should be also confronted with the supernova theory.

More generally, the gravitational waves discovery, GW150914, may reveal the new population of weak gamma ray bursts originating from the low Lorentz factor jets. In this case, the prompt electromagnetic emission may be difficult to discover, but the observation of a radioactive decay of elements produced in the engine of the GRB may be confronted with the engine simulation predictions. The black hole parameters, namely its spin and mass, determine the shape of the gravitational wave signal and hence can be independently constrained. As shown in Figure \ref{fig:nse2Dm62}, the results for nucleosynthesis yields somewhat depend on the black hole spin value. The abundance of isotopes produced in the
accreting torus around a faster spinning black hole, is about 10 times higher
at about $A=80$ (Selenium group) than for a slowly spinning black hole. Thus, the predictive power of our models is worth exploring 
in the future work, if only the accreting 
black hole parameters are determined and the nucleosynthesis yields constrained with some future observations, i.e. X-ray probes.

\section*{Acknowledgments}

We thank Michal Bejger, Szymon Charzynski, Bartek Kaminski, and Petra Sukova 
for helpful discussions, and Irek Janiuk for help in the MPI parallelization of the HARM-2D
code.
We also thank the anonymous referee for constructive comments which helped
us to improve the presentation of our results.
This research was supported in part by grant DEC-2012/05/E/ST9/03914
from the Polish National Science Center.
We also acknowledge support from the Interdisciplinary Center for Mathematical Modeling of the 
Warsaw University, through the computational grants G53-5 and gb66-3.

\end{document}